\documentclass[twocolumn,prl,superscriptaddress,floatfix]{revtex4}
\usepackage{graphicx}
\usepackage{xcolor}
\usepackage{epsfig}
\usepackage{rotating}
\usepackage{amsmath}
\usepackage{amsfonts}
\usepackage{amssymb}
\usepackage{enumerate}
\usepackage{longtable}
\usepackage{appendix}
\usepackage{dcolumn}
\usepackage{bm}
\DeclareMathOperator{\Tr}{Tr}
\usepackage{hyperref}
\usepackage{ulem}

\begin{document}

\title{Finite Temperature Strong Coupling Expansions for the SU(N) Hubbard Model}

\author{Rajiv R. P. Singh}
\affiliation{Department of Physics, University of California Davis, 
CA 95616, USA}

\author{Jaan Oitmaa}
\affiliation{School of Physics, The University of New South Wales,
Sydney 2052, Australia}

\date{\rm\today}

\begin{abstract}

We develop finite temperature strong coupling expansions for the SU(N) Hubbard Model in powers of $\beta t$, $w=\exp{(-\beta U)}$
and ${1\over \beta U}$ for arbitrary filling. The expansions are done in the grand canonical ensemble and are most useful at a density of one particle per site, where for $U$ larger than or of order the Bandwidth, the expansions converge over a wide temperature range $t^2/U \ \lesssim \ T \ \lesssim \ 10 U$. By taking the limit $w\to 0$, valid at temperatures much less than $U$, the expansions turn into a high temperature expansion for a dressed SU(N) Heisenberg model that includes nearest-neighbor exchange, further neighbor exchanges and ring exchanges known from the $T=0$ perturbation theory of the SU(2) Hubbard model. Below a filling of one particle per site, the $w\to 0$ limit corresponds to an effective $t-J$ model. The onset of strong correlations can be identified by a plateau-like behavior in the entropy as a function of temperature.
At small deviations from one particle per site, the expansions can be arranged in powers of a small parameter $\delta=1-n$, the deviation from one particle per site, where the leading $\beta t$ dependent terms correspond to holes sloshing around in a disordered SU(N) background.
We use these expansions to calculate the thermodynamic properties of the model at moderate and high temperatures over a wide parameter range.

\end{abstract}


\maketitle

\section{Introduction}

Recent developments in the physics of cold atoms has allowed substantial progress to be made in our understanding of the Fermi Hubbard model \cite{greiner,brown,mitra,cheuk,rosch,gross}, one of the most important models in Condensed Matter Physics. The ability to artificially synthesize representations of the Hubbard model with well characterized interaction parameters, combined with new types of experimental measurements, some of which are impossible in the solid state, has given the field a huge boost and an opportunity to think about many old and new aspects of equilibrium and non-equilibrium many-body phenomena.

Another recent development is the study of cold-atom systems which offer a generalization of the well studied SU(2) Hubbard model with two spin species to the SU(N) Hubbard model with N species of Fermions \cite{bloch,honerkamp,taie,padilla}. Taking these Fermi Hubbard systems down to very low temperatures remains a challenge for experiments, but already interesting behavior can be seen at moderate to high temperatures. 

The purpose of this paper is to develop systematic finite-temperature strong coupling expansions for the SU(N) Hubbard model. The expansions are developed in the thermodynamic limit in the grand canonical ensemble as a function of the fugacity $\zeta$ in powers of $\beta t$, $w=\exp{(-\beta U)}$ and ${1\over \beta U}$. The expansion coefficients are simple polynomials in the SU(N) parameter $N$. 

The expansions are most useful at one particle per site, where they converge over a wide temperature range. By taking the limit $w\to 0$, the expansions turn into a high temperature expansion for a generalized SU(N) Heisenberg model that contains nearest- and further-neighbor exchange interactions as well as multi-spin exchange interactions. In second order of the expansions, we identify the nearest-neighbor exchange interaction, which is order $t^2/U$. In fourth order perturbation theory we identify higher order terms of order $t^4/U^3,$ that includes nearest-neighbor exchange, second-neighbor exchanges between sites that share a neighbor and 4-spin processes in a ring. For the case of $N=2$, our results agree completely with the earlier work of MacDonald, Girvin and Yoshioka \cite{macdonald,delannoy}. Our calculations give us the SU(N) generalization of these parameters. There is a change in sign of the ring exchange terms as a function of N. 

In second order perturbation theory, the entropy at low temperatures saturates at the value of $\ln{N}$, as only the constant, non-zero trace terms in the effective Hamiltonian, contribute to the partition function in this order. But in fourth-order, spin correlations begin to develop due to nearest-neighbor exchange interactions of order $t^2/U$. They lead to a reduction in entropy within the single-occupancy subspace. Sixth order of the expansions give deviations of the entropy function from the nearest-neighbor Heisenberg model. We will present numerical results for various properties of the square-lattice SU(N) Hubbard model for $N=2$, $3$ and $4$ for moderate to large values of $U$. These expansions are not particularly useful numerically for small $U$, as $U$ goes in the denominator. 

Moving away from one-particle per site, we focus on particle densities, $\rho$, less than one-particle per site.
We present numerical results for the square-lattice SU(N) Hubbard models at various densities, where one can see a crossover from the high temperature regime ($T>U$) to a strongly correlated regime
at temperatures  $T<<U$. The strongly correlated regime is characterized by a plateau-like behavior in the entropy function, which is nearly $U$ independent for large $U$. In the $w\to 0$ limit, these expansions can be turned into a high temperature expansion for a generalized SU(N) $t-J$ model. 
Close to one particle per site, the hopping terms are small by a parameter $\delta=1-n$. The leading $\delta$ dependent terms correspond to isolated holes sloshing around in a disordered SU(N) background. 

We believe our calculation should serve as a benchmark for other numerical calculations and for cold atom experiments on SU(N) systems with moderate to large $U$ values especially at intermediate ($T\simeq t$) and high ($T>>t$) temperatures.

\section{Model and Methods} 

The SU(N) Hubbard model is defined by a Hamiltonian $H=H_0+V$, where the unperturbed Hamiltonian $H_0$ is an on-site term:
\begin{equation}
   H_0= U \sum_i {n_i (n_i-1)\over 2} -\mu \sum_i n_i,
\end{equation}
with $n_i$ the total number operator for particles on site $i$ and $\mu$ is the chemical potential. The perturbation $V$ is the hopping term:
\begin{equation}
    V=-t\sum_{<i,j>}\sum_{\alpha=1}^N (C_{i,\alpha}^\dagger C_{j,\alpha} + h.c.),
\end{equation}
where the sum $<i,j>$ runs over nearest-neighbor pairs of sites of a lattice and the sum over $\alpha$ runs over the $N$ species of Fermions.

Using the formalism of thermodynamic perturbation theory \cite{oitmaa-book, oitmaa2},the logarithm of the grand partition function, per site, can be expended as
\begin{equation} \label{Eq_pert}
    \begin{split}
    &\frac{1}{N_s} \ln{Z} = \ln{z_0} + \\
    &\sum_{r=1}^\infty \int_0^\beta d\tau_1 
    \int_0^{\tau_1} d\tau_2 \ldots \int_0^{\tau_{r-1}} d\tau_r <\Tilde{V}(\tau_1)\ldots\Tilde{V}(\tau_r)>_N
    \end{split}
\end{equation}
where $z_0$ is the single-site partition function, $N_s$ is number of sites in a large system,
\begin{equation}
    \Tilde{V} =e^{\tau H_0} V e^{-\tau H_0},
\end{equation}
and,
\begin{equation}
    <X> = \Tr{e^{-\beta H_0}X}/\Tr{e^{-\beta H_0}}.
\end{equation}
 Let $\zeta$ be the fugacity defined as $\zeta=e^{\beta \mu}.$ The particle density (per site) can be obtained via the relation
\begin{equation}
    \rho=\frac{\zeta}{N_s}{\partial \over \partial \zeta} \ln{Z}.
\end{equation}
This relation needs to be solved to obtain $\zeta$ or $\mu$ as a function of $\rho$ and $\beta$, which then allows one to obtain various properties at fixed particle density.

We can readily obtain other thermodynamic quantities such as Internal energy $E$ and entropy $S$ using the relations
\begin{equation}
    E = -({\partial \over \partial \beta} \ln{Z})_\zeta,
\label{E}
\end{equation}
and
\begin{equation}
    S=-\beta({\partial \over \partial \beta}\ln{Z})_\zeta -\rho \ln{\zeta} + \ln{Z}.
    \label{S}
\end{equation}
We define two measures of double occupancy as
\begin{equation}
    D_1=-\frac{1}{\beta}{\partial \over \partial U}{\ln{Z}\over N_s},
    \label{D1e}
\end{equation}
and 
\begin{equation}
    D_2=w { \partial \over \partial w}{\ln{Z}\over N_s},
\label{D2e}
\end{equation}
where in $D_2$ no differentiation is performed with respect to the $1/{\beta U}$ terms.
Note that $D_2$ will go to zero when $w\to 0$ but $D_1$ will not as it includes all virtual double occupancy terms. Vanishing $D_2$ gives us the onset of the strongly correlated regime or generalized Heisenberg or $t-J$ model.

\section{Series Expansions}

The single-site partition function is a series in powers of $w$ and is readily obtained as
\begin{equation}
\begin{split}
    z_0=&1+N\zeta + {N(N-1)\over 2}\zeta^2 w \\
    &+ {N(N-1)(N-2)\over 6}\zeta^3 w^3 + \mathcal{O}(w^6),
    \end{split}
\end{equation}
Since we are interested in moderate to large $U$ and at temperatures below $U$, the higher order terms are exponentially small and can be safely neglected.

\begin{figure}[hbpt]
\includegraphics[width=1\columnwidth]{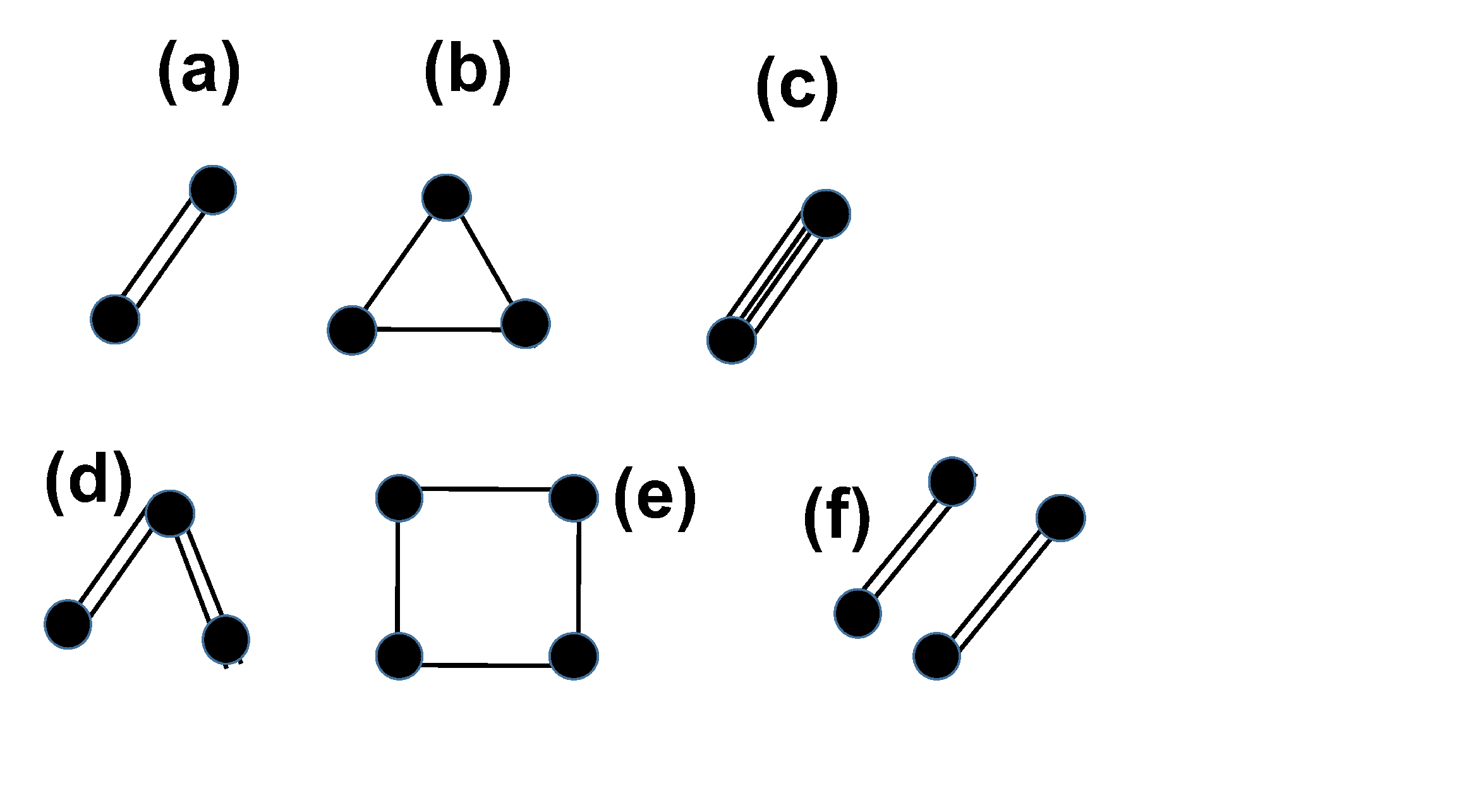}
\caption{Graphs that contribute to fourth order of the expansion. Graph (a) contributes in second order, Graph (b) in third order and graphs (c), (d), (e) and (f) in the fourth order. }
\label{graphs}
\end{figure}

In each order, the terms in the expansion can be expressed in terms of various graphs on the lattice as:
\begin{equation}
    \frac{1}{N_s}\ln{Z}=\ln{z_0}+\sum_{G}L_G z_0^{-s} (\beta t)^r X_G(\zeta,\beta U,N),
\end{equation}
where the sum is over graphs denoted $G$. The graph $G$ has $s$ sites and $r$ bonds. $L_G$ is the lattice constant of the graph defined as the extensive part of the graph count, per lattice site. The weight-factor $X_G(\zeta,\beta U, N)$ is the reduced contribution of the graph obtained from an evaluation of the trace. The graphs that contribute up to fourth order are shown in Fig.~\ref{graphs}. The contributions from each graph rapidly become quite cumbersome and need to be evaluated by a computer program. Note that in this formalism, disconnected graphs are also needed. More details on the method can be found in Oitmaa {\it et al} \cite{oitmaa-book}. 

The second order term comes from just a single graph shown as (a) in Fig.~\ref{graphs}. It has two sites and two powers of the same bond connecting the sites give us the second order in $\beta t$. The trace calculations lead to the weight for the first graph $X_a$ equal to:
\begin{equation}
\begin{split}
    &N(\zeta+(N-1)^2\zeta^3w+\frac{1}{4}(N-1)^2(N-2)^2\zeta^5w^4) \\
    &+2N(N-1)\zeta^2(1+\frac{1}{2}(N-1)(N-2)\zeta^2w^2)(1-w)/\beta U \\
    &+ \frac{1}{2}N(N-1)(N-2)\zeta^3w(1-w^2)/\beta U \\
    &+\frac{1}{6}N(N-1)^2(N-2)(N-3)\zeta^5w^4(1-w^2)/\beta U\\
    &+\frac{1}{9}N(N-1)(N-2)(N-3)\zeta^4w^3(1-w^3)/\beta U+\ldots,
    \end{split}
    \label{X2}
\end{equation}
where the neglected terms represented by $\ldots$ are order $w^6$ and hence numerically extremely small at the temperatures below $T=U$. As a reminder, our interest is in the temperature regime $t^2/U< T < 10 U$ with $U$ of order the bandwidth or larger.

There is only one graph in 3rd order consisting of a triangle of 3 bonds. This graph is absent on bipartite lattices. In fourth order there are 3 connected graphs and one disconnected graph. The expressions rapidly become too unwieldy for use without a computer program.

Since the fourth order terms become important only at temperatures of order $t$ and by this temperature $w$ becomes exponentially small, it suffices to focus on the $w\to 0$ limit for the fourth and higher order terms. Note that this limit does not imply we are considering a strict large-U limit, as we have all inverse power of $1/U$ still present in the calculations. In this $w\to 0$ limit, the largest power of $\zeta$ in a graph is given by the number of sites in the system. Furthermore, for any given power of $\zeta$ the $N$-dependence is a polynomial whose order is the power of zeta. Thus, at fourth order, the $N$ dependence of these terms is fully determined from knowing the results up to $N=4$. 

For the graphs labelled by the letters (c), (d) and (e), let us call the weight-factors $X_c$, $X_d$ and $X_e$ respectively. In the $w\to 0$ limit, these simplify considerably to become
\begin{equation}
    X_c=\frac{N\zeta}{12} + 4N(N-1)\frac{\zeta^2}{(\beta U)^2}-8N(N-1)\frac{\zeta^2}{(\beta U)^3}
    \label{Xc}
\end{equation}
\begin{eqnarray}
 X_d=&{N\zeta\over 6} + {N^2\zeta^2\over 6}+
        {3N(N-1)\zeta^2 \over \beta U}
    - {4N(N-1)\zeta^2 \over (\beta U)^2} \nonumber \\
   & +{4N(N-1)^2\zeta^3 \over (\beta U)^2}
   +{2N(N-1)\zeta^2 \over (\beta U)^3} \nonumber\\ &+{N(N-1)(10N-8)\zeta^3 \over 3(\beta U)^3}
   \label{Xd}
\end{eqnarray}
and
\begin{eqnarray}
 X_e=&{N(\zeta-4\zeta^2+\zeta^3)\over 3}
 +{4N(N-1)(\zeta^2-\zeta^3)\over \beta U} \nonumber \\
 &+ {4N(N-1)\zeta^2(-2-5\zeta+3N\zeta)\over 
 (\beta U)^2} \nonumber \\
 &+{8N(N-1)\zeta^2(1 +(5-3N)\zeta +(5-5N+N^2)\zeta^2 )  \over (\beta U)^3}
 \label{Xe}
\end{eqnarray}
The weight-factor of a disconnected graph such as graph (f) in Fig~\ref{graphs} is the product of the weights of its disconnected pieces.

We should note that there is a subtlety in taking $w\to 0$ limit at one particle per site $\rho=1$ in that $\zeta$ becomes exponentially large as $1/\sqrt{w}$ (see next section). However, any additional power of $\zeta$ relative to one particle per site always brings with it an additional power of $w$, so that these terms remain exponentially small relative to the leading terms.

\section{One particle per site}
The expansions simplify greatly when the particle density $\rho=1$. At low temperatures ($T<t$), the chemical potential can be determined up to exponentially small correction terms and in the $w\to 0$ limit the system maps into a generalized SU(N) Heisenberg model. In this section, we focus on this mapping analytically. Numerical results for different N will be presented at the end of the section.

The single-site partition function, keeping only the lowest power of $w$, is
\begin{equation}
    z_0 = 1 + N\zeta +{N(N-1)\over 2}\zeta^2 w.
\end{equation}
Thus,
\begin{equation}
    \rho=\frac{1}{z_0} (N\zeta + N(N-1)\zeta^2 w).
\end{equation}
Setting $\rho=1$ leads to the result
\begin{equation}
    \zeta^2 =\frac{1}{w} {2\over N(N-1)}.
\end{equation}
This relation between chemical potential and U is exact for $N=2$ at all temperatures but not so for larger N as noted in the work by Padilla et al \cite{padilla}. However, it allows a systematic expansion for all higher order terms where exponentially small terms in powers of $\sqrt{w}$ can be neglected.
We obtain,
\begin{equation}
    z_0=2 + N\zeta.
\end{equation}
Hence, in the $w\to 0$ limit, we have the relation:
\begin{equation}
    \frac{\zeta}{z_0}=\frac{1}{N} + \mathcal{O}(\sqrt{w})
\end{equation}
Since the $w\to 0$ limit of $X_G$ for a graph with $s$ sites has $z_0^s$ in the denominator and the numerator has a maximum power of $\zeta^s$ without any double occupancy only the $\zeta^s$ terms survive in this limit. All terms which bring additional powers of $\zeta$ bring additional powers of $w$ as well and thus remain exponentially small. Thus, we obtain (focusing on the terms relevant to the square-lattice)
\begin{equation}
    \frac{X_a}{z_0^2}=\frac{2(N-1)}{N}\frac{1}{ \beta U},
\end{equation}

\begin{equation}
    \frac{X_c}{z_0^2}= \frac{4(N-1)}{N}\frac{1}{(\beta U)^2}-\frac{8(N-1)}{N}\frac{1}{(\beta U)^3},
\end{equation}
\begin{equation}
 \frac{X_d}{z_0^3}=
   {4(N-1)^2 \over N^2(\beta U)^2}
   +{(N-1)(10N-8) \over 3N^2(\beta U)^3},
\end{equation}
and
\begin{equation}
 \frac{X_e}{z_0^4}=
 {8(N-1)(5-5N+N^2)   \over N^3(\beta U)^3}.
\end{equation}
It is interesting no note that the last term, which is related to ring exchanges changes sign between $N=3$ and $N=4$.

The terms of order $\beta$ in the logarithm of the partition function correspond to $<-\beta H_{eff}>$ that is minus the trace of different effective Hamiltonian terms obtained in a $t/U$ expansion of the Hubbard model \cite{macdonald,delannoy}. For example the leading term for $N=2$ is $\beta t^2/U$ or $\beta J/4$, which corresponds to the $-J/4$ term in the well known second order effective Hamiltonian for the Hubbard model at half filling
\begin{equation}
    J(\vec S_i \cdot \vec S_j - 1/4)
\end{equation}
The terms $\beta t^4/U^3$ come from one bond, two bond and 4-site ring graph. For $N=2$, these become $-4\beta t^4/U^3$, $\beta t^4/U^3$ and $-\beta t^4/U^3$. These agree with the constant terms in the work of MacDonald {\it et al} \cite{macdonald} and set the magnitudes for various exchange parameters of the effective model.

Thus, finite temperature perturbation theory is an alternative method for determining the exchange parameters in the effective Hamiltonian. However, any linear term in $\beta$ in $\ln{Z}$ does not contribute to reduction in the entropy. Thus, in second-order perturbation theory, the entropy at low temperatures saturates to $\ln{N}$, corresponding to the singly occupied subspace. We need to go to the fourth-order terms to see the reduction in entropy within the singly occupied subspace. In fourth-order the reduction in entropy comes from the $\beta^2 (t^2/U)^2$ terms. Thus, for large $U$ this gives us the leading nearest-neighbor spin correlations from Heisenberg interactions. This term is equivalent to the Curie-Weiss or mean-field behavior of the Heisenberg model. To see additional correlations due to higher order terms in $t/U$  
one would have to go to sixth and even higher orders of perturbation theory. We will discuss these higher order corrections to the entropy at the end of this section.

Combining contributions from all the graphs to fourth order in perturbation theory and multiplying their contributions by the lattice constants for the square-lattice which are $2$, $2$, $6$, $1$ and $-7$ for graphs (a), (c), (d), (e) and (f) respectively, and adding them up, our results for the logarithm of the partition function per site becomes:
\begin{equation}
    {\ln{Z}\over N_s}= \ln{z_0} + \frac{\beta t^2}{U} a_N + \frac{\beta^2 t^4}{U^2} b_N +  \frac{\beta t^4}{U^3} c_N,
\end{equation}
where
\begin{equation}
    a_N={4(N-1) \over N},
\end{equation}
\begin{equation}
    b_N={8(N-1) \over N} -{4(N-1)^2\over N^2},
\end{equation}
and,
\begin{eqnarray}
    c_N=&{-16(N-1) \over N} +{4(N-1)(5N-4)\over N^2} \nonumber \\
    &+{8(N-1)(5-5N+N^2)\over N^3}.
\end{eqnarray}
For the $N=2$ the first two terms add up to ${\beta J\over 2}+\frac{3}{16}\beta^2J^2$, with $J=4t^2/U$. The $\beta^2$ term agrees with the known results for the Heisenberg model \cite{glenister}. The first two coefficients $a_N$ and $b_N$ depend on $(1-1/N)$ only. We believe, this is the source of `Universality' observed by Padilla {\it et al}
\cite{padilla}. The last term is smaller by two powers of $t/U$.
These are also the first terms absent in second order Numerical Linked Cluster Expansions \cite{rigol}, invoked in the study of Padilla et al \cite{padilla}.

The entropy function, per site, on the square-lattice to fourth order becomes
\begin{equation}
    \frac{S}{N_s} = \ln{N} -b_n \frac{\beta^2 t^4}{U^2}
\end{equation}
with the $\beta^2$ dependence  characteristic of all such lattice models.

\begin{figure}[htb!]
\centering
\includegraphics[angle=270, width=\columnwidth]{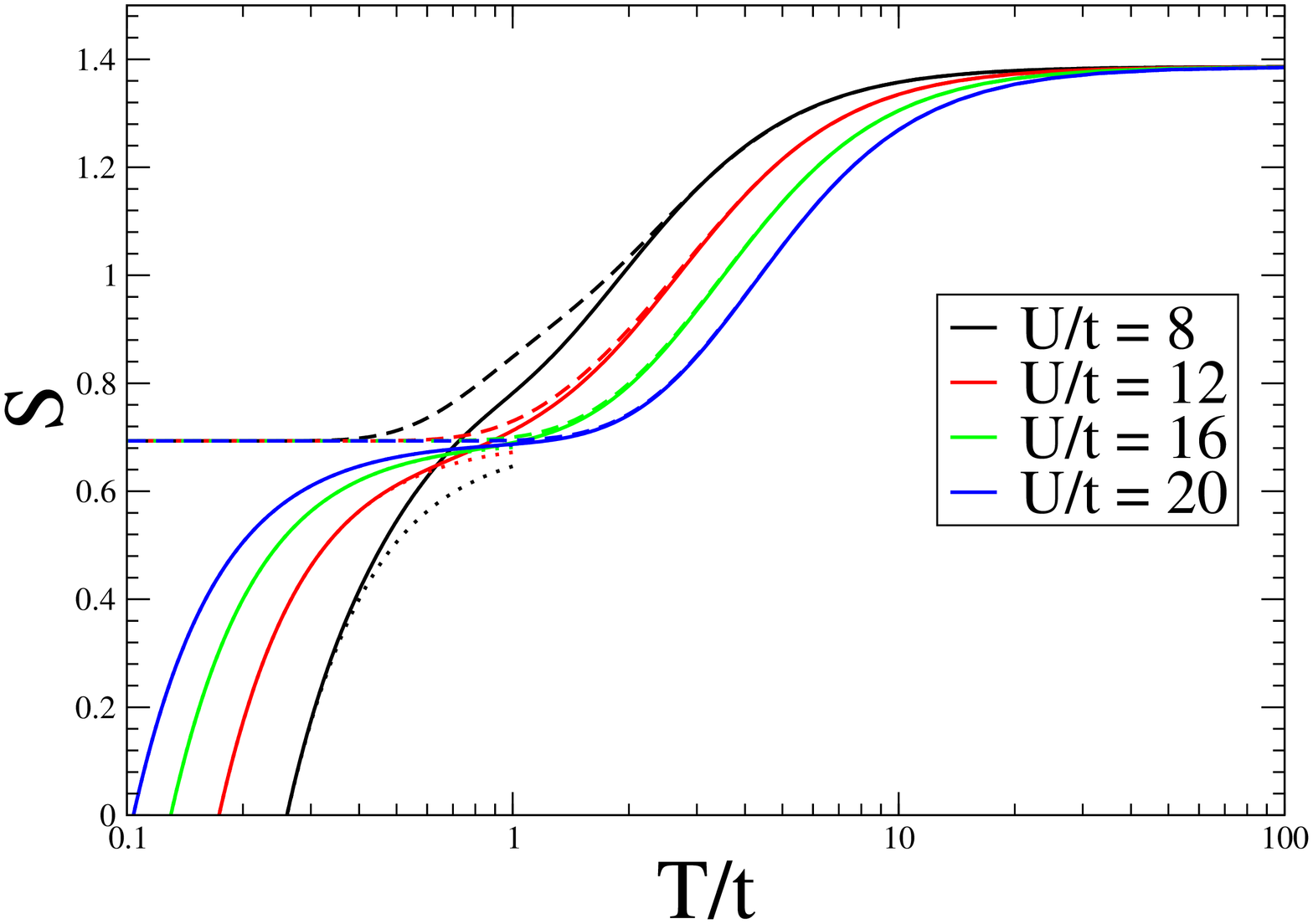}
\includegraphics[angle=270, width=\columnwidth]{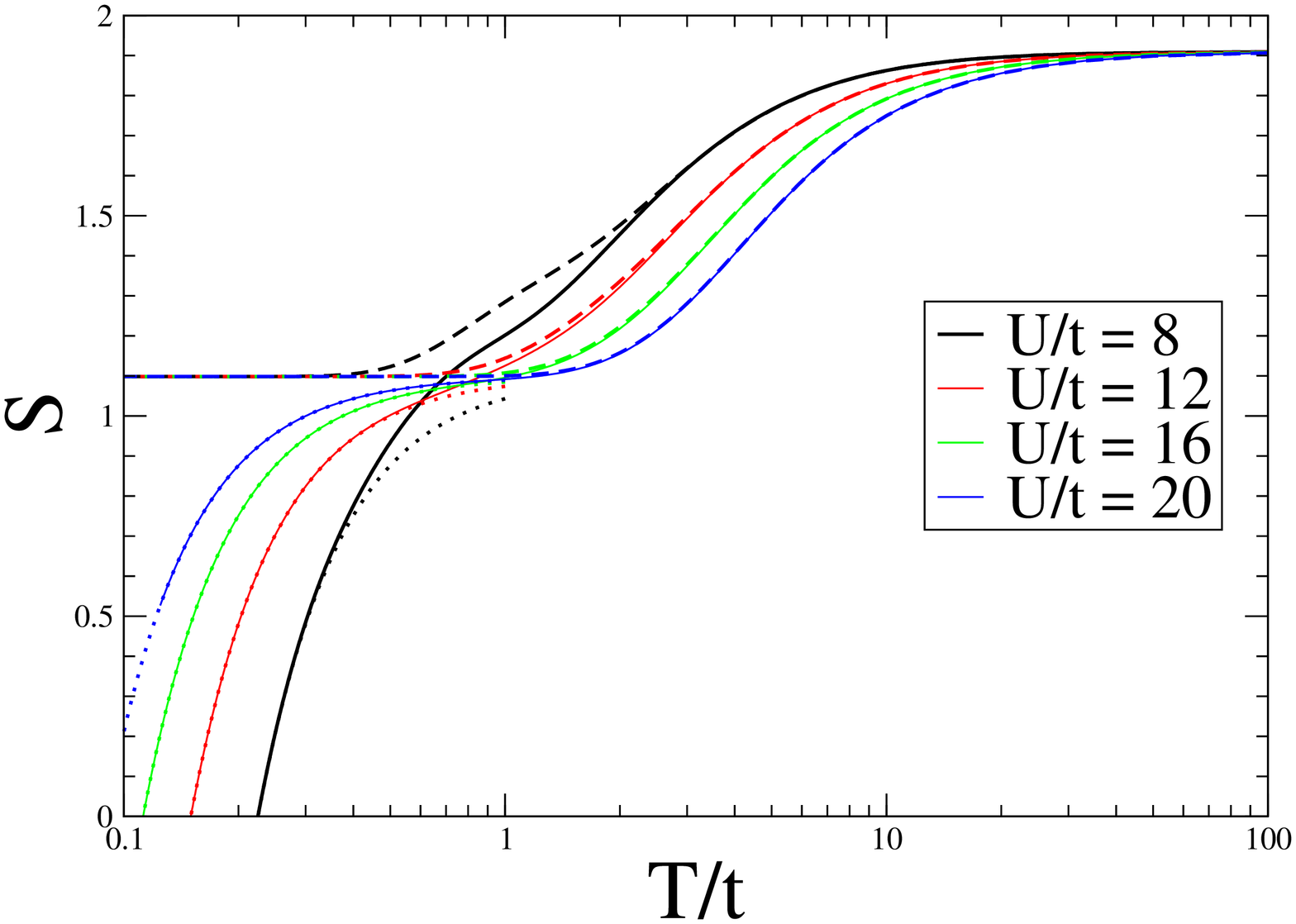}
\includegraphics[angle=270, width=\columnwidth]{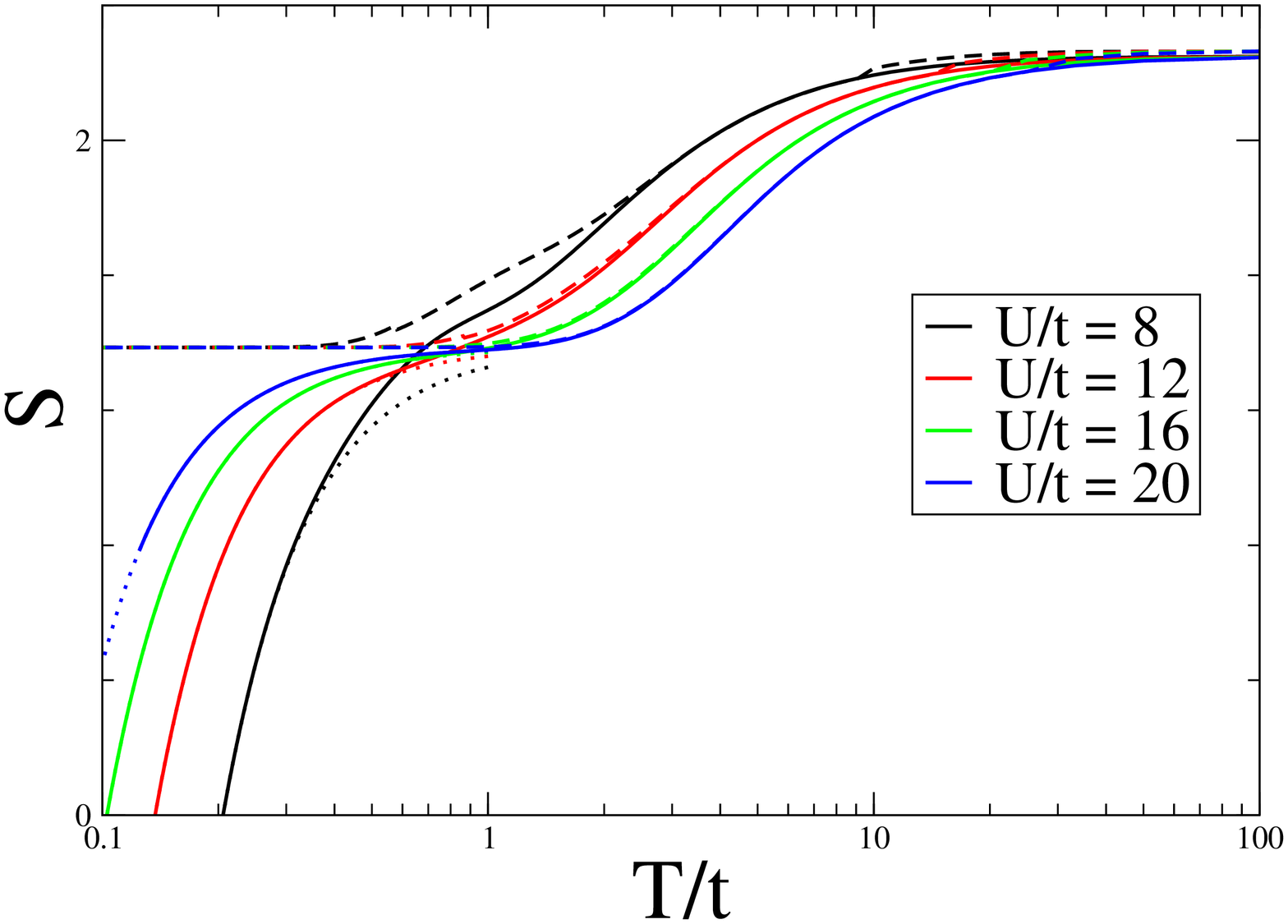}
\caption{Entropy as a function of temperature for N=2 (upper plot), N=3 (middle plot) and N=4 (lower plot). For each value of N and $U$ the full fourth order perturbation theory result is shown by a solid line, the second order perturbation theory result is shown by a dashed line and the $w\to 0$ limit, generalized Heisenberg limit is shown by a dotted line.}
\label{entropy} 
\end{figure}

In Fig.~\ref{entropy}, we show plots of entropy per site $S$, for
several $U$ values for $N=2$, $N=3$, and $N=4$. Results are shown from the full second and fourth order perturbation theory as well as from the reduced fourth order perturbation theory which is valid at temperatures much less than $U$ and provides a mapping to the generalized Heisenberg model. 

One can see that the generalized Heisenberg model works quite well at temperatures below $t$ for all $U$. Note that our generalized Heisenberg model includes ring exchange terms. For $U=8$, the temperature scale for the applicability of this model is such that by that time the entropy is significantly below $\ln{N}$. There is no real entropy plateau at $\ln{N}$ for $U/t=8$. This is an important result for experiments, where it is common to fit changes in entropy between low and high temperatures to $\ln{2}$ for a spin-half system. Our calculations show that this is only valid if $U/t$ is greater than about $10$. For smaller $U$ values, there is no temperature window where the double occupancy can be ignored and the system still has nearly the full $\ln{N}$ entropy left.

\begin{figure}[htb!]
\centering
\includegraphics[angle=270, width=\columnwidth]{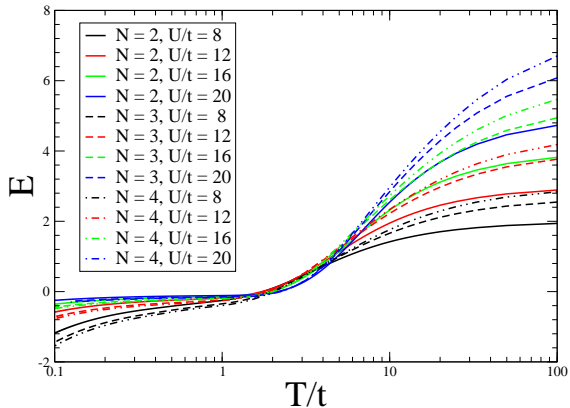}
\caption{Energy obtained from the full fourth order perturbation theory is shown for various N and $U$. }
\label{energy} 
\end{figure}

\begin{figure}[htb!]
\centering
\includegraphics[angle=270, width=\columnwidth]{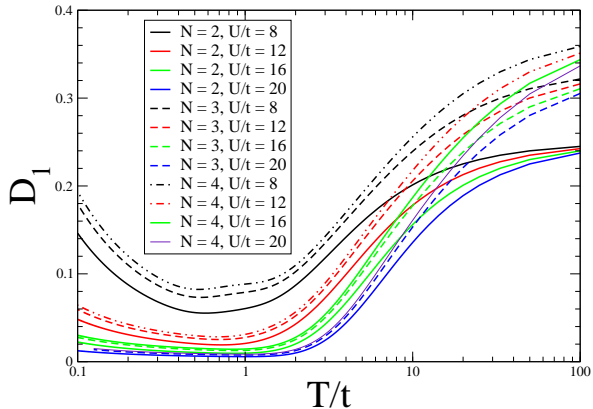}
\caption{Double Occupation Measure $D_1$ obtained from the full fourth order perturbation theory is shown for various N and $U$. }
\label{D1} 
\end{figure}

\begin{figure}[htb!]
\centering
\includegraphics[angle=270, width=\columnwidth]{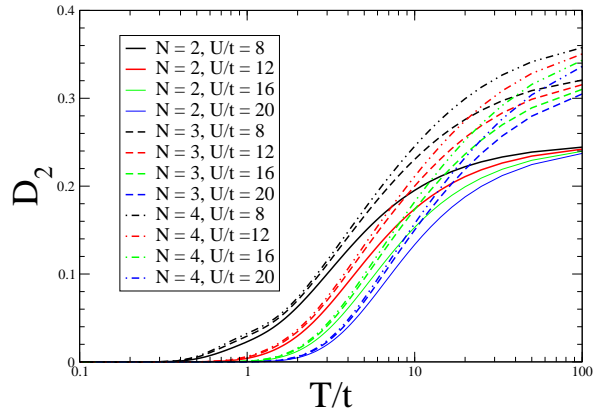}
\caption{Double Occupation Measure $D_2$ obtained from the full fourth order perturbation theory is shown for various N and $U$. Note that this is defined such that it must go to zero as the exponentially small terms with positive powers of $w$ go to zero. }
\label{D2} 
\end{figure}

In Fig.~\ref{energy}, we show the energy function obtained from the full evaluation of the fourth order perturbation theory. In Fig~\ref{D1} and Fig.~\ref{D2}, the double occupancy is shown as a function of temperature. As noted earlier $D_1$ is a true measure of double occupancy while $D_2$ must go to zero when $w\to 0$ and it marks the onset of the effective generalized Heisenberg model.

Since, the fourth order terms already converge well for $T<t$, for higher orders of perturbation theory, one can confine one selves to the $w\to 0$ limit. This greatly simplifies the trace calculations and makes it possible to do the sixth order calculations. We have
carried out the sixth order calculations for $N=2$, $3$ and $4$. For the SU(2) case, the sixth order expansions for $\ln{Z}$ are:
\begin{equation}
    2\ \beta^3 \ \frac{t^6}{U^3} - 36 \ \beta^2 \ \frac{t^6}{U^4} + 62.5 \ \beta \ \frac{t^6}{U^5}.
\end{equation}
The first term is just the $\beta^3$ term for the Heisenberg model and agrees with known results \cite{glenister}. The third term comes from traces of higher order ($t^6/U^5$) new terms in the effective Hamiltonian generated by perturbation theory. The second term is in some sense the most interesting. It leads to the leading deviations of the entropy 
function from the nearest-neighbor Heisenberg model. We find that for the SU(2) case this deviation in the entropy in second order in $\beta$ is only $3$ percent for $U/t=20$ but rises to close to $20$ percent for $U/t=8$. This further shows that by $U/t=8$ higher order terms in $t/U$ cannot be ignored in any quantitative description of the system.

For SU(3) case, the sixth order expansions for $\ln{Z}$ are
\begin{equation}
    \frac{128}{81}\ \beta^3 \ \frac{t^6}{U^3} - \frac{2048}{81} \ \beta^2\  \frac{t^6}{U^4} + \frac{175904}{2187}\  \beta \ \frac{t^6}{U^5},
\end{equation}
And, for SU(4) case, the sixth order expansions for $\ln{Z}$ are
\begin{equation}
    \frac{5}{4}\  \beta^3\ \frac{t^6}{U^3} - \frac{45}{4} \ \beta^2 \ \frac{t^6}{U^4} + \frac{50789}{576} \ \beta \ \frac{t^6}{U^5}.
\end{equation}

\section{Particle densities below one particle per site}
Although our expansions are valid for arbitrary particle densities, we will focus here on particle densities less than one per site ($\rho<1$). At these densities, the strong correlation limit, or $w\to 0$ limit, corresponds to an effective SU(N) t-J model. However, we now have hopping terms that scale as $\beta t$ and these terms begin to grow at a temperature below $T=t$. This limits the convergence of our expansions numerically to the temperature range $T>t$. It is known that a high order series expansion converges reasonably for the SU(2) $t-J$ model at all densities down to much lower temperatures \cite{putikka,glenister,pryadko}. But, keeping the higher powers of $t/U$ coming from the Hubbard model \cite{mila} and developing a high temperature expansion for an extended t-J model has, to our knowledge, not been done even for the SU(2) case, and we leave this for future work.

From a numerical point of view, there is a simplification when densities are close to one particle per site. In this case one can expand properties in powers of $\delta=1-\rho$. For a single site, the partition function (setting $w=0$) becomes
\begin{equation}
    z_0 = 1 + N\zeta.
\end{equation}
The bare density is given by
\begin{equation}
    \rho_0={N\zeta \over 1 + N\zeta}.
\end{equation}
This can be inverted to give
\begin{equation}
    N\zeta = {\rho_0 \over 1-\rho_0},
\end{equation}
and, thus,
\begin{equation}
    z_0 = {1\over \delta_0},
\end{equation}
where $\delta_0=1-\rho_0$ serves as a small parameter. The grand partition function in zeroth order becomes
\begin{equation}
    \frac{\ln{Z}}{N_s} = -\ln{\delta_0},
\end{equation}
But, because $z_0$ goes in the denominator, all higher order terms, including the full density, can be expanded in positive powers of $\delta_0$. In leading order $\delta=1-\rho=\delta_0.$
Hopping terms are small by a factor of $\delta_0$.
However, the leading order term in $\delta_0$ gets contributions from all powers of $\beta t$ and it corresponds to isolated holes in a fluctuating SU(N) background, which is an important problem in itself \cite{trugman}.

It is straightforward to expand expressions for $X_a$ through $X_e$ given earlier in powers of $\delta_0$ to obtain the  free energy of the doped system. We only give here the leading order result. To order $t^2$ and $\delta_0$, one obtains:
\begin{equation}
    \frac{\ln{Z}}{N_s} = -\ln{\delta_0} + 2\rho \beta^2t^2 \delta_0 +\frac{4(N-1)}{N} \frac{\beta t^2}{U}(1-2\delta_0).
\end{equation}
The second term comes from the hopping of an isolated hole. Since hopping Hamiltonian is traceless, it contributes first only to second order in the logarithm of the partition function. The last term is just the Heisenberg term reduced due to doping. 
The asymptotic high temperature expression for the t-J model becomes
\begin{equation}
    \frac{S}{N_s}=-\delta\ln{\delta} -\rho\ln{\rho} +\rho\ln{N}-2\rho \beta^2t^2 \delta,
    \label{asymp}
\end{equation}
where $\delta=1-\rho$.

\begin{figure}[htb!]
\centering
\includegraphics[width=\columnwidth]{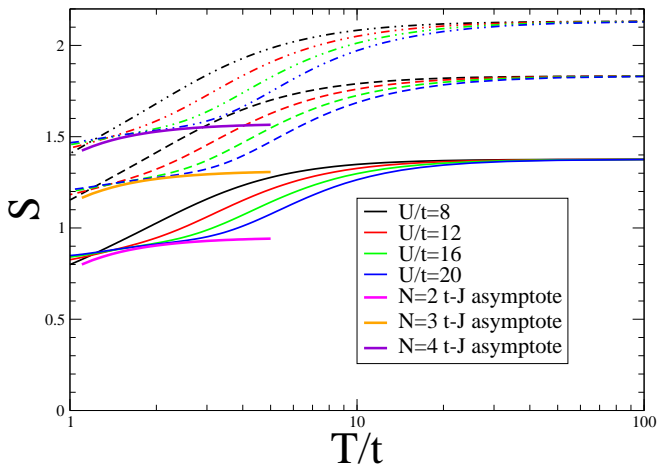}
\caption{Entropy function for $\rho=0.9$ obtained from the full fourth order perturbation theory is shown for various N and $U$ values. The $N=2$ cases are shown by solid lines, $N=3$ cases by dashed lines and $N=4$ cases by dash-dotted lines. The color is same for a given value of $U$.}
\label{rhoa} 
\end{figure}

\begin{figure}[htb!]
\centering
\includegraphics[width=\columnwidth]{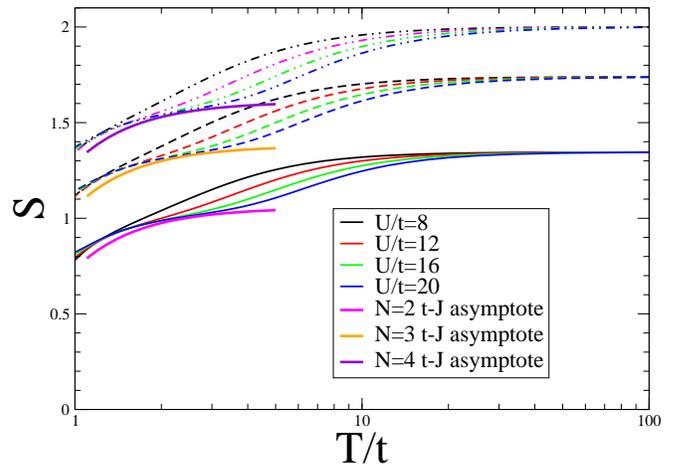}
\caption{Entropy function for $\rho=0.8$ obtained from the full fourth order perturbation theory is shown for various N and $U$ values. The $N=2$ cases are shown by solid lines, $N=3$ cases by dashed lines and $N=4$ cases by dash-dotted lines. The color is same for a given value of $U$.}
\label{rhob} 
\end{figure}

In Fig.~\ref{rhoa} and Fig.~\ref{rhob}, we show plots of the entropy function at intermediate to high temperatures for densities of $\rho=0.9$ and $0.8$ respectively, and several $U$ values of the order of  or larger than the bandwidth for $N=2$, $3$, $4$ from a complete numerical evaluation of all the weights to fourth order. Plots for the same $U$ values are shown by the same color. The results for different $N$ values are qualitatively similar. Except for the case of $U=8$, all the larger $U$ values show a clear intermediate temperature regime. There is a crossover from the high temperature regime at temperatures of order or larger than $U$ to a strongly correlated regime at temperatures well below $U$. The $U=8$ case, is at the boundary of developing such a regime.

This intermediate temperature regime is characterized by a plateau-like flattening of the entropy curves, which becomes more and more pronounced as $U$ increases. Furthermore the entropy function in this regime becomes nearly $U$ independent. This indicates that double occupancy is strongly reduced at these temperatures ($w\to 0$) and it marks the onset of an effective $t-J$ model. The asymptotic high temperature expression for the entropy of the t-J model
\ref{asymp} is shown in the plots by the solid lines. The energy scale is at first dominated by $t$ only as the temperatures are high enough that the effects of $J$ are still negligible. Thus, the entropy becomes nearly $U$ independent. the exchange $J$ plays an important role in further reduction of entropy below this plateau-like region. Hence, going to temperatures below $T=t$ the drop is most pronounced for smaller $U$ values. This intermediate temperature regime and crossover could be explored in cold atom experiments.


\section{Discussions and Conclusions}
In this paper, we have developed a finite temperature perturbation theory for the SU(N) Hubbard model for arbitrary fugacity $\zeta=e^{\beta\mu}$, in powers of $\beta t$, $w=\exp{-\beta U}$ and $1/(\beta U)$. The expansions at second order are complete for all $N$ up to $\mathcal{O} (w^6)$. Up to fourth order, terms are complete for $N=2$, $3$ and $4$. For higher $N$ we have the full $N$ dependence when $w\to 0$ that is when the temperature is much less than $U$. For $N=2$, $3$ and $4$, we have also obtained the sixth order expansions in the $w\to 0$ limit. These expansions are particularly useful when $U$ is of order or larger than the bandwidth and there is one-particle per site. In this case these expansions are well convergent over a wide temperature range $t^2/U \lesssim T \lesssim 10 U$, {\it i.e.} from a high temperature many times $U$ down to a temperature of order the exchange constant or the mean-field ordering temperature. 

At low temperatures ($w\to 0$ limit), these expansions for the Hubbard model turn into a high temperature expansion for a generalized Heisenberg model, which contains nearest and further neighbor spin exchanges as well as ring exchange terms and other high order processes. In second order of perturbation theory, the entropy saturates at low temperatures to the value of $\ln{N}$, the full entropy associated with the one-particle subspace. Fourth and higher order terms contain development of spin correlations and consequent further reductions in entropy of the system within the singly occupied subspace.

For particle density per site $\rho$ less than unity, these expansions turn into a high temperature expansion for a generalized $t-J$ model when $w\to 0$. Because the expansions now contain various powers of $\beta t$, without any inverse powers of $U$, the naive convergence of the expansion is set by $T=t$ rather than $T=t^2/U$. High order series expansions \cite{putikka,glenister,pryadko} are needed to go down to temperatures of order and below $J$. 

In the temperature regime $T>t$, where the expansions converge well for $\rho$ less than unity, one can already see a crossover to a strongly correlated regime ($T<<U$) from a high temperature regime ($T$ of order or larger than $U$). This regime is characterized by a flattening of the entropy function as a function of temperature. Furthermore, the entropy function becomes nearly $U$ independent for large $U$ as the exchange energy scale $J$ is still negligible at these temperatures. The study of this crossover may be accessible to cold atom experiments.

We hope our work will serve as a benchmark for other numerical studies as well as for experiments on cold atom systems.
The expansions can also be used to study other Mott phases at integer $\rho$ values for $N$ greater than 2. These will be discussed in a future work.

{Acknowledgement:} This work is supported in part by the US National Science Foundation grant DMR-1855111.

\end{document}